\documentclass{article}

\usepackage{arxiv}

\usepackage[utf8]{inputenc} % allow utf-8 input
\usepackage[T1]{fontenc}    % use 8-bit T1 fonts
\usepackage{hyperref}       % hyperlinks
\usepackage{url}            % simple URL typesetting
\usepackage{booktabs}       % professional-quality tables
\usepackage{amsfonts}       % blackboard math symbols
\usepackage{nicefrac}       % compact symbols for 1/2, etc.
\usepackage{microtype}      % microtypography
\usepackage{graphicx}
\usepackage{booktabs} % For formal tables

\usepackage[ruled]{algorithm2e}
\usepackage{lineno,hyperref}
\usepackage{fancyhdr,amssymb}
\usepackage[inline]{enumitem}
\usepackage{array}
\usepackage{caption}
\usepackage{subcaption}
\usepackage{amsmath}
\usepackage{amsthm}
\usepackage{pbox}
\usepackage{lipsum}
\usepackage{mathtools}
\usepackage{cuted}

\title{Hybrid-Learning approach toward situation recognition and handling}

\author{
  Hossein Rajaby Faghihi\\
  Department of Computer Engineering\\
  Sharif University of Technology\\
  Tehran\\
  \texttt{faghihi@ce.sharif.edu} \\
   \And
 MohammadAmin Fazli \\
  Department of Computer Engineering\\
  Sharif University of Technology\\
  Tehran\\
  \And
  Jafar Habibi \\
  Department of Computer Engineering\\
  Sharif University of Technology\\
  Tehran\\
}

\begin{document}
\maketitle

\begin{abstract}
The success of smart environments largely depends on their smartness of understanding the environments' ongoing situations. Accordingly, this task is an essence to smart environment central processors. Obtaining knowledge from the environment is often through sensors, and the response to a particular circumstance is offered by actuators.
This can be improved by getting user feedback, and capturing environmental changes. Machine learning techniques and semantic reasoning tools are widely used in this area to accomplish the goal of interpretation. In this paper, we have proposed a hybrid approach utilizing both machine learning and semantic reasoning tools to derive a better understanding from sensors. This method uses situation templates jointly with a decision tree to adapt the system knowledge to the environment. To test this approach we have used a simulation process which has resulted in a better precision for detecting situations in an ongoing environment involving living agents while capturing its dynamic nature.
\end{abstract}

% keywords can be removed

\section{Introduction}
\label{sec:introduction}
The boost of techniques in machine learning and pervasive computing coincide with the increased accessibility of environmental data have caused an increasing interest in developing smart environments. Moreover,  smart environments have shown a promised influence on a variety of areas such as healthcare monitoring, elderly assisting, surveillance, and so forth.
Although the research has begun from novel papers,  soon it is going to play an essential role in daily human life due to the reduction of the power consumption of the electronic components, and the spread of the wireless communication technologies \cite{mukhopadhyay2014internet}. 

As the sensors of smart environments increased in number and improved in robustness, an opportunity arose due to the availability of the data generated by sensors; The data is useful in understanding the ongoing situation in the environment, and to offer relevant services to the environment's users. The data generated from sensors are almost meaningless unless there is a data analysis or interpretation. The act of analyzing the raw data generated by sensors emerge the opportunity of the actual understanding of smart environments by central processors handling them.

Although the phase of interpretation is essential to acquire actual knowledge about smart environments, it is quite challenging due to several characteristics of an environment which involves living agents.
The first exquisite characteristic of such environments is their dynamic nature. These environments are not likely to follow a particular pattern during a long period and tend to change frequently. The occurring changes can be both in the behavior of living agents or the availability and reliability of sensors.
Another challenge of smart environments is the huge amount of data generated frequently by plenty of sensors through time to be interpreted.
Moreover, The data itself is often not quite reliable due to the corruption of communication tools or the accuracy of the sensors themselves.
Furthermore, processing of the recent data received from a sensor is often inadequate to understand changes in an environment. Hence, storing historical data is a mandatory task. For instance, we can not determine if a man has entered a room just by watching his presence in the room. We certainly require his previous location to identify if she has just entered the room or she was already present there.

Situation-Awareness is widely mentioned in the area of HCI\footnote{Human-Computer Interaction} \cite{matheus2003core}. Hence, considering the most critical goal of situation-aware systems, which is helping the user to achieve its objective by offering relevant services, situation awareness is not attained unless the human relationship with the environment is captured.

%The dynamic nature of an environment which involves living agents is because of the user behavior changes and possible element movements. Initially, the living agents may change their habits of acquiring a goal. Furthermore, the elements of an environment may be moved and their location would change.

The process of understanding the smart environment circumstances follows
\begin{enumerate*}
	\item gathering data from sensor networks
	\item deriving more abstract elements from sensor data 
	\item identifying occurring situations by considering the abstract elements provided by the second step.
\end{enumerate*} \cite{d2015multi}.
The abstract data derived from the sensor data is commonly referred to as context data.
Context is defined by Dey and Abowd as any information that can be used to characterize the situation of an entity, where the entity is an object playing a role in the environment \cite{dey2001understanding}. For instance, when a sensor sends a data element of type integer equal to 20, and that sensor type is a thermostat, the context would get that number and relate it to the temperature of the location in which that thermostat sensor is resided. The results of interpretation would be that the location ${x}$ temperature is 20 centigrade for the timestamp ${y}$ and that's what we call context information.
The ability of a smart environment to be aware of context information is called context-awareness. Context-Awareness is defined as the use of context by the smart environment to provide task-relevant information or services to a user \cite{abowd1999towards}. Context-awareness is considered mainly as a key factor to develop intelligent and practical systems to function in the most important today concerns. This is considered important in many criteria such as intelligent transportation \cite{al2013context, wan2014context, santa2009sharing, ferreira2018context, rehena2018towards al2018qos}, smart cities \cite{al2018small, urbieta2017adaptive} and healthcare \cite{abdellatif2019edge, aborokbah2018adaptive, casino2018smart, al2018context}.

As mentioned in the process, the context data would have to be interpreted again by accumulating all the contexts in a process named situation identification. Situations as described in \cite{kolbe2017enriching} are a more complex structure than context; it is mentioned that situations involve complex relations and temporal aspects with a need for additional semantic interpretation.
When the smart environment becomes aware of its situations, the smart environment is called situation-aware. Situation-Awareness is defined as the understanding of the elements in an environment regarding of time and space, the appreciation of their semantic, and a hypothetical status of what is going to happen to them \cite{endsley1995measurement}.

%What contexts contain can help us know about different phenomena separately in the real world, but is it adequate to help the person acting in the environment to accomplish his tasks?. The answer would be negative for many occasions. For instance, consider you are going to heat the water in a pot to drink some tea, being aware that your location is near the pot, or the water temperature is getting higher or even that you are hosting some guests doesn't help the smart environment to understand the fact that you are going to have some tea by considering  these pieces of knowledge separately, but all this context information together may do.

Many methods and approaches have been proposed for achieving a situation-aware intelligent system. There exist methods which use learning while others rely on semantic and specification-based methods \cite{ye2012situation}. Newly approaches combine methods mentioned before and solve this issue with a hybrid method though. Hybrid methods tend to use the benefits of both approaches while eliminating their weaknesses \cite{ye2012situation}.

This paper introduces a new hybrid approach toward situation identification consistent with the works presented in \cite{hirmer2017situation}. We show that our method has a superior  in accuracy, flexibility and adaptability in comparison with the base research which enables a to have better accuracy in dynamic environments.

In the remainder of this paper,we review previous related works in Section \ref{sec:relatedwork}. In Section \ref{sec:biologicalreview} challenges, and issues considered in order to make such changes into situation identification process are discussed. The method is presented in Section \ref{sec:proposed_model} and eventually the evaluation of the proposed method of identification is in Section \ref{sec:evaluation}.

\section{Background and Related Works}
\label{sec:relatedwork}
As described in the previous section, by and large, there are three main categories of approaches toward situation identification.
The first relies on machine learning techniques.
This kind of methods generally considers situations as labels, and sensors' data as inputs \cite{yang2016towards}. However, the input data in situation recognition techniques are massive and require streaking pre-processing steps. Some common algorithms used in this area are Neural Network \cite{yang2008using}, Naive Bayes \cite{korpipaa2003bayesian}, Decision Tree \cite{alkhomsan2017situation, lee2016situation,logan2007long}, hidden Markov models \cite{lee2016situation,damarla2008hidden}, support vector machine \cite{kanda2008will}, Bayesian networks \cite{ye2012situation, korpipaa2003bayesian} and Genetic Algorithms \cite{yang2016towards}.
All these methods share the problem of lacking expert knowledge and cold start.

%Decision Tree is a function in which inputs are coming from one side in and labels are going out from another side relating to that input. 
%Decision Tree has several implementations like ID3\cite{quinlan1986induction} and C4.5\cite{michie1993quinlan}. 
%This kind of tree generally split the input data according to one feature in each level of tree and each node, and finally, when reaching one leaf node, returns the most occurring label as leaf node label in training mode and then after that, the tree would use the construction built with training data in order to find the path from root to one leaf for each new input data.

The second category of approaches toward situation identification are specification-based algorithms.
This kind of methods generally relies on logical reasoning and rule engines\cite{ye2012situation}.
One subcategory is logic programming \cite{barwise1989situation}. Utilizing logic programming for situation-awareness is through specifying each occurring situation as a set of rules and logic sets. The limitations in this approach are originated from the constraints of description logic.
Another technique is to use ontology. Ontology is defined as explicit specification for a concept \cite{gruber1993translation}. Hence, ontology defines each concept by its definition and relations to other concepts.
The main focus in this technique is to design an ontology regarding the problem domain. The knowledge in such systems comes from the understanding of its ontology designer who tried to formalize the knowledge in concept definitions and relations. By using ontology, the knowledge represented would be shareable, transferable, and understandable by both machines and humans concurrently \cite{chen2009semantic}. Ontology is a popular approach toward situation identification in many papers \cite{ghimire2017iot, matheus2003core, pearson2016generic, kolbe2017enriching, machado2017reactive}. Situation template is another subcategory of specification-based approaches. Situation templates are one directional XML trees with a root labeled as the situation and specifying the situation model by using XML \cite{mormul2017situation, hirmer2017situation, hirmer2015sitrs, da2016sitrs}. 
Spatial and temporal logic are also used as a side technique or the main technique. This is because identifying situations without the knowledge of the location and time of the events is rarely possible in many cases \cite{cook2009ambient}. 
Another research topic in this area is fuzzy ontology or fuzzy logic. These methods are widely used to control uncertainty in sensor data or process steps \cite{ranganathan2004reasoning, ke2018automatic}.

This research is mainly based on a framework proposed by Hirmer et al \cite{hirmer2017situation} which is called "SitOPT". SitOPT has presented a three-layer framework containing sensor layer, situation recognition layer, and workflow handling layer. The sensor layer handles communication with physical objects and system hardware. Its duty is to gather sensor data and translate them into context information, and then pass them to the situation recognition layer. The situation recognition layer is responsible for identifying the occurring situations according to received input from the sensor layer, and finally, the situation identified will be transferred to the workflow layer where it is translated into a set of tasks and then be executed to affect the environment through actuators.

Situation handling in the SitOPT framework is based on Situation templates \cite{hirmer2015sitrs}. 
As mentioned before, situation templates are data flow structures similar to trees. Each node of these trees is either an input sensor variable or a constant value at the leaf level.
Upper nodes are some operators comparing two or more of the leaves or other operator nodes, and finally, the root is a situation node with a name attached to it. When there is a path from leaves to the root node that every equation in nodes holds (comparing operators to the values from the input), the situation specified in the root is recognized as an occurring situation. The tree structure is derived from an older work, Situation-Aggregation-Trees (SAT) \cite{zweigle2009supervised}.

"SitOPT" is incapable of capturing the dynamic nature of environments. Its situation recognition method is an specific-based algorithm that inherits common problems of these approaches described in Section \ref{sec:introduction}. The SitOPT deployment is suggested to be done with the Node-RED platform which executes data flows in clouds, local, and any other devices \cite{hirmer2017situation}.

Matheus et al \cite{matheus2003core} develop a core ontology for situation-aware applications. They claimed that although it is important to know about objects, the knowledge about objects' relationships are what makes situation identification possible. They introduce a core ontology for situation-awareness indicating that every situation-aware application should customize the ontology according to its specific situations. In the proposed ontology, situations are related to goals, rules and physical objects.

\begin{figure}[!h]
	\centering
	\includegraphics[scale=0.5]{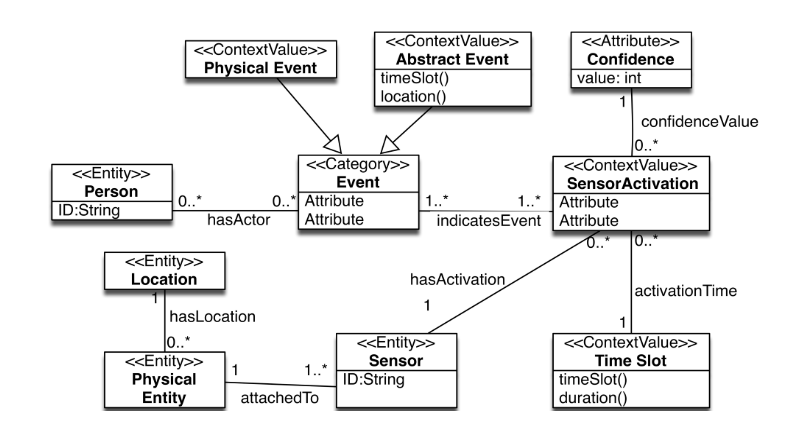}
	\caption{Core ontology for context modeling}
	\label{fig:context}
\end{figure}

In our research ontology is not used in the main method. However, we use ontology for interpreting sensor data into context. Cause SitOPT method didn't directly introduce any transformation from sensor data to context information, we use an ontology to understand the meaning of sensor data in the scope of an ontology core \cite{pearson2016generic}. This ontology should be extended to each environment by experts before installing the intelligent system.
The research of Pearson et al \cite{pearson2016generic} has also used ontology to achieve situation-awareness. They developed two separate ontologies, one for context modeling and one for situation modeling. The context ontology is presented in Figure \ref{fig:context}. This ontology describes an event with the use of the sensor alerting it, the location of the sensor, the timestamp that the event is happening in and finally, the reliability in sensor value by the use of a confidence concept.

Hybrid approaches are a more recent point of view in order to understand the ongoing situation by combining both specific-based methods and machine learning methods. Yuan and Herbert \cite{yuan2014context} introduced a mixed learning algorithm in healthcare scope as a context-aware real-time assistant using rule-based and case-based reasoning together.
Using machine learning techniques to generate dynamic association rules and adding the sound rules to adapt ontology is another method to enable hybrid-learning \cite{kim2017augmented}.

In this paper, we present a hybrid-learning algorithm based on situation templates and decision tree combining as online and offline reasoning tools. Because of the specific structure of decision trees, they are easily integrated with situation templates. Additionally, unlike other machine learning technique which will not specify how a label is resulted from inputs, the decision tree rules for decision making is clear and human-understandable.

\section{Brief Overview of Challenges}
\label{sec:biologicalreview}
To get the best results from the process of situation recognition, an essence is to understand the environmental characteristics and design accordingly.
These characteristics are the best reasons to design a hybrid-learning algorithm instead of a specification-based algorithm. The most important factors are discussed in the followings.

The dynamic nature of active environments is already discussed in Section \ref{sec:introduction}. This is our first driver to enable self-adaptation in the situation recognition system.

Another difficulty one may encounter when interacting with smart environments is data corruption. Data corruption can either be a consequence of a broken sensor or the result of uncertainty in sensor values. Uncertainty indicates the fact that the accuracy of the data received from a sensor shouldn't be 100\% trusted \cite{alberti2013internet}. Regarding this fact, we have to design the processes enabling them to handle errors and to prevent incorrect reasoning as the consequence of erroneous inputs. Additionally, broken sensors will result in either a shortage of required data input for interpretation or a misleading input values.

Another issue to consider is the huge amount of data flowing from sensors to the central processor unit. Sensors are producing values frequently, and these generated data are transformed and compiled into either context information or situations when received by the central processor unit. The processor doesn't necessarily know which data is useful for the identification process leading the system to face a significant amount of data for the interpretation \cite{sukode2015context}. If a smart system has to store every image of environment sensors data or to process the whole input data at every data arrival, it would need a high number of resources and a significant amount of time to extract any meaningful data or retrieve an old stored data from warehouses. Considering the described issue jointly with the necessity of keeping historical information to recognize changes or patterns in an environment, which indicates the urgency of storing data into warehouses, would enlighten us why IoT applications face a big data problem if not applying any filtering mechanism on data processing.

Another fact to consider is that in every human-computer interaction system there exist humans! If every task is done in an automated manner or if the user doesn't play a role in the decision cycle, then it is not fitted into actual user needs. When designing a smart environment, what must be considered critical is the people living there and the way they interact with their surroundings. The user must feel so that the whole system is under his control and its mission is to help her to do daily tasks more easy and quick. This type of interaction requires two primary considerations. First, the user must be aware of the system procedures and be able to change what she desires. Second, the system must adapt itself to the user needs and help her suffer less in completing a routine or daily task.

Our Solution to all the challenges mentioned above will be entirely discussed in the Section \ref{sec:proposed_model}. However, there is a brief description in the followings.

First of all, regarding the issue of the dynamic nature of environments, this paper has considered the solution of dynamic knowledge through machine learning techniques to adapt the situation models to the dynamics.
Secondly, the solution to uncertainty is divided into two steps in this paper. Initially, the approach considers the accuracy of sensor data as a relation between sensor data and confidence node in the context model mentioned in Figure \ref{fig:context}. Furthermore, it has used machine learning techniques to detect broken sensors and remove them from the situation models.

Thirdly, the solution to the challenge of big data is filtering. As we have used the situation template in the identification process, filtering would be automatically applied as situation templates just consider predefined data types instead of evaluating all the inputs provided to them.

In the end, the solution to user interaction and involvement is applied through situation manipulation and adaption. This paper personalizes occurring situations by learning from user feedback without specifying exact changes in situation templates to the user. This is done by using human labeled data as inputs to a machine learning technique. As a result, the user doesn't have to understand all the complexities of the situation templates' structure, and changes would be more reliable because the machine learning method only changes the templates if a pattern is frequent.

\section{The Proposed Hybrid Approach}
\label{sec:proposed_model}
\begin{figure}[!h]
	\centering
	\includegraphics[scale=0.12]{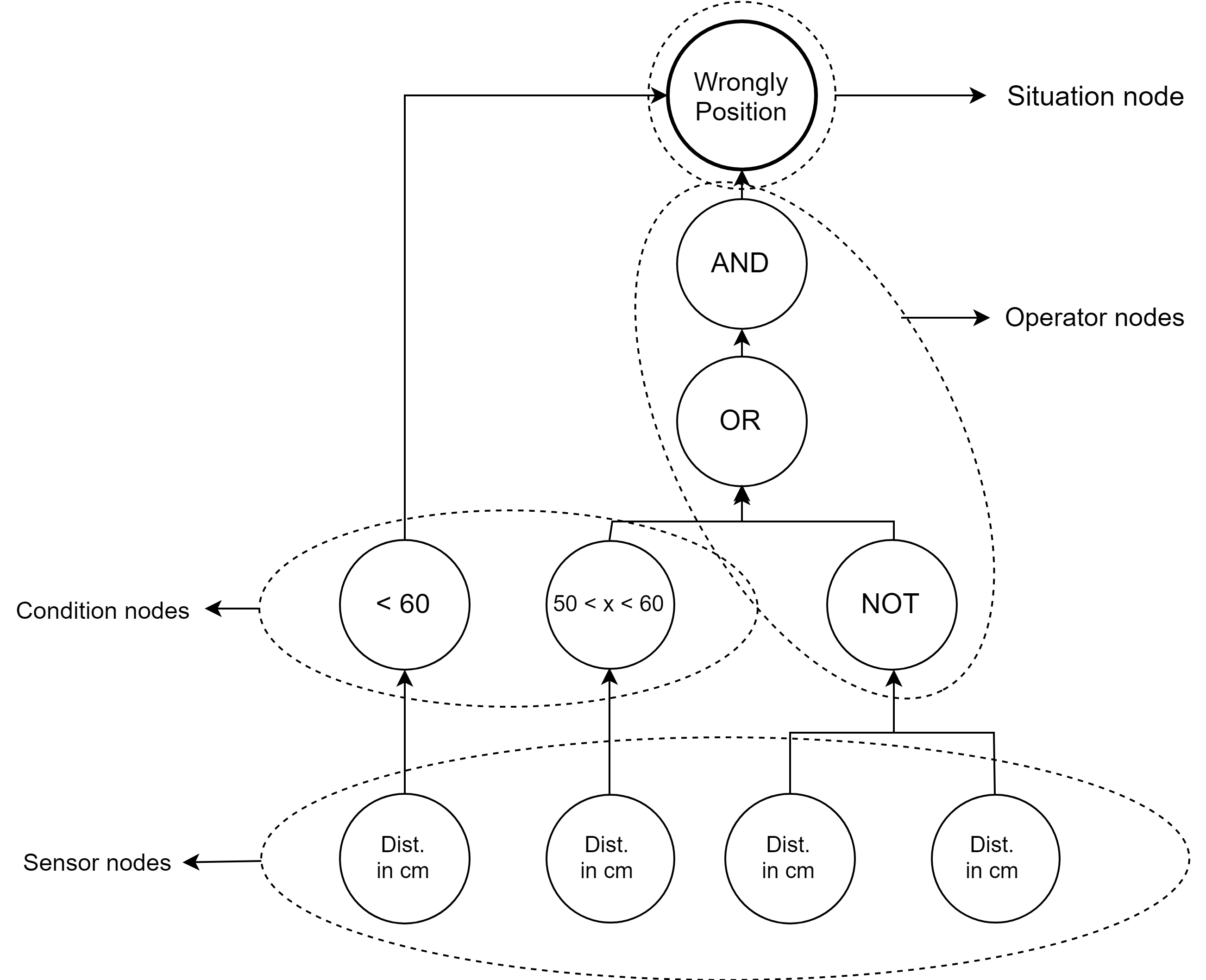}
	\caption{A situation template sample}
	\label{fig:sitrs}
\end{figure}

The proposed hybrid approach gives a solution to the challenges mentioned in Section \ref{sec:biologicalreview} and will be compatible with SitOPT framework discussed in Section \ref{sec:relatedwork}. As explained in Section \ref{sec:relatedwork}, this paper will improve the SitOPT's situation recognition layer so it becomes more accurate and adaptable regarding changes. 
SitOPT's situation modeling is based on situation templates. These templates are tree-like models made up of operator, condition, and sensor nodes jointly with a single situation node which are shown in Figure \ref{fig:sitrs}.

As situation templates are built by human experts, 
\begin{enumerate*}
	\item Lack of environment understanding by the experts,
	\item No consideration of sensor corruption,
	\item No consideration of environment's living agents behavior,
\end{enumerate*}
are three possible fault origins that can lead to erroneous models. Moreover, \begin{enumerate*}
	\item New added sensor,
	\item Sensor removal,
	\item Human behavior change,
\end{enumerate*}
are some other factors that indicate the need for an adaption strategy to modify the situation models.

\begin{figure}[!h]
	\centering
	\includegraphics[scale=0.15]{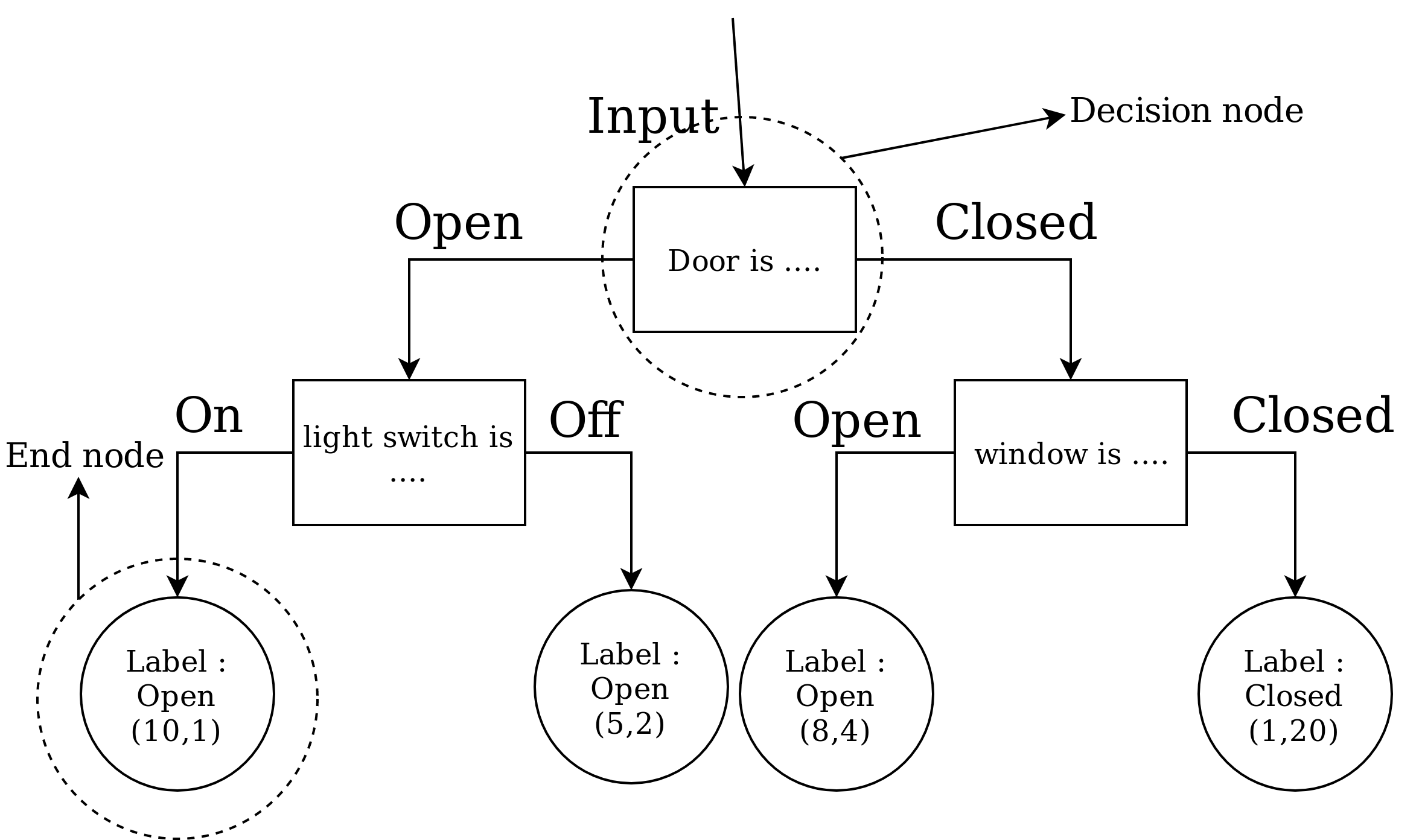}
	\caption{A decision tree sample}
	\label{fig:dt}
\end{figure}

As described in Section \ref{sec:relatedwork}, a decision tree \cite{safavian1991survey} is used to enable adaptation in the situation models in our approach. A decision tree is a decision support tool that uses a tree-like model to correspond labels with inputs. This tree is consisted of decision and end nodes. Each decision node splits the data according to their features' vector and each ending node is a leaf which assigns a label to them. A sample decision tree is shown in Figure \ref{fig:dt}.

Decision trees and situation templates are very resembled. However, a decision tree is in a reverse direction of what a situation template is. Accordingly, labels in situation templates are on the roots while the leaves represent labels in a decision tree. 

\begin{figure*}[!h]
	\centering
	\hspace*{-2cm}
	\includegraphics[scale=0.18]{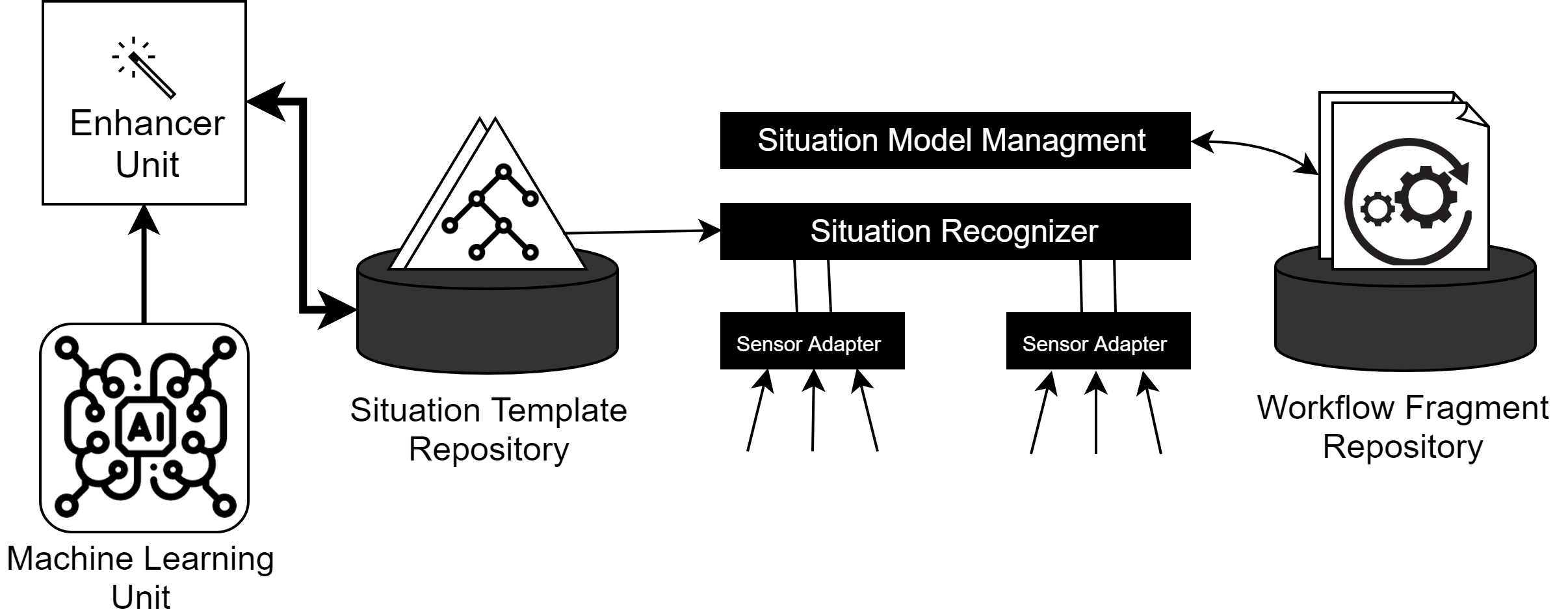}
	\caption{The structure overall view}
	\label{fig:structure}
\end{figure*}

The situation identification unit's structure of our approach is shown in Figure \ref{fig:structure}. In this structure, the machine learning unit is responsible for capturing sensors' images jointly with user feedback to generate a decision tree. It works in relation to the situation repository where all situation models are stored as situation templates. Moreover, an enhancer unit is responsible for merging the decision tree with the situation templates in the repository. Executing the situations and recognizing what situation corresponds to the sensor's images received from the sensor layer, is the recognizer unit goal. Furthermore, the workflow fragment repository is where the situations are mapped into environmental tasks or reflections. The workflow fragment repository, the situation repository, and the recognizer unit have already been present in the architecture of SitOPT.

As stated before, the machine learning unit works by a decision tree algorithm. Decision trees use a variety of algorithms to decide which feature of the input vector should be chosen as a decision node at each level of the tree. We use C4.5 algorithm \cite{quinlan1993c4} in our approach. What enables us to use a machine learning unit which requires labeled inputs is the fact that almost all tasks in a smart environment which involves living agents should be semi-automated or at least implemented with an alerting tactic. This way, the user feedback is collected jointly with the sensors' image which satisfies the labeled inputs requirement. After the tree is built by the training data, each tree's end node comes with a purity and a cardinality measure. The cardinality measure indicates the number of training data concluded in an end node. Moreover, The purity metric is computed by dividing the cardinality of the inputs with the prevalent label in the node to its cardinality.

\begin{figure*}[!h]
	\centering
	\hspace*{-1cm}
	\includegraphics[scale=0.15]{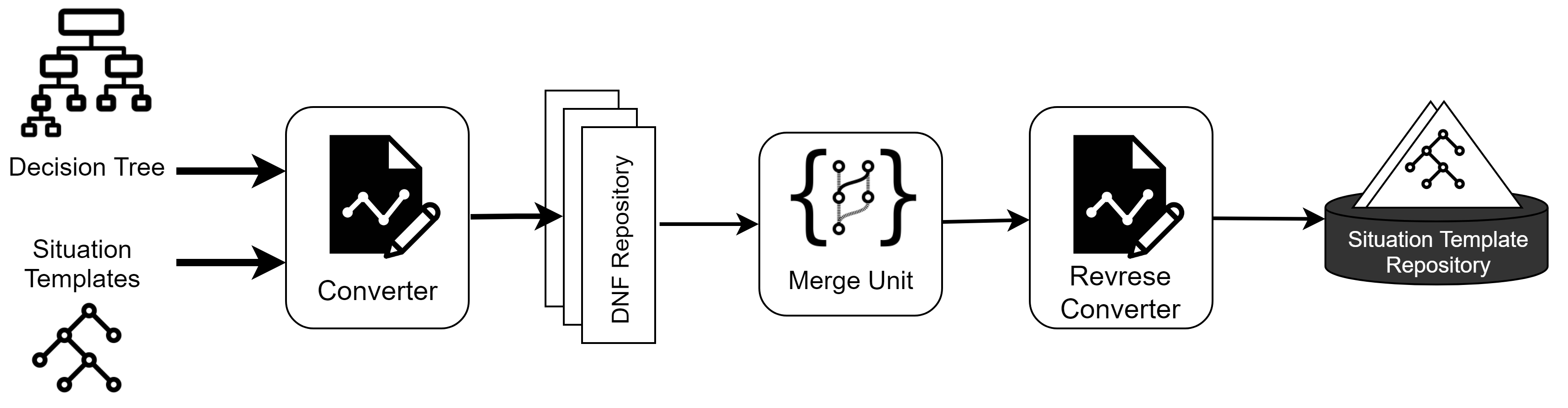}
	\caption{The enhancer process overview}
	\label{fig:process}
\end{figure*}

The process which is run in the enhancer unit is shown in Figure \ref{fig:process}. Initially, the process works by converting both the decision tree and the situation templates into the similar structure known as DNF. After the conversion process, each label (situation) would be described in two DNF trees; The first tree is resulted from the decision tree paths concluding in an end node holding that label, and the second tree is corresponded to the equivalent situation template in the situation repository.

\begin{figure}[!h]
	\centering
	\includegraphics[scale=0.16]{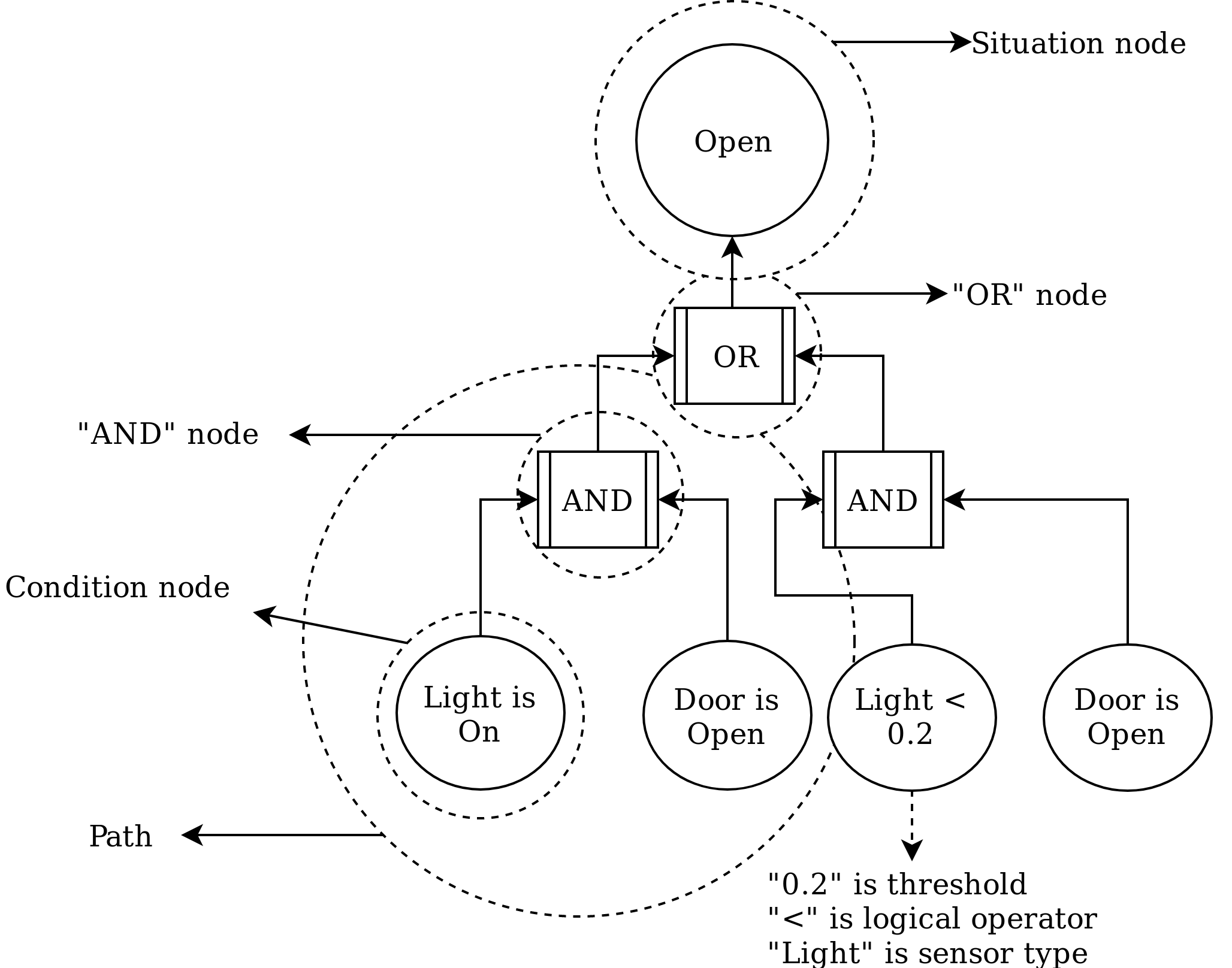}
	\caption{DNF tree}
	\label{fig:dnf}
\end{figure}

A sample output of a DNF tree is shown in Figure \ref{fig:dnf}. DNF trees in our method are made up of condition, "AND" operator nodes, a single "OR" operator node and a single situation node as the root.  A path in these trees is a sub-tree starting with an "AND" operator node. These paths determine different circumstances which a situation can be occurring based on a sensors' image. As each branch of the decision tree correspond to one path of a  DNF tree, each path would inherit the purity and cardinality measure.
After converting both machine learning output and human knowledge into DNF trees, the merging process begins. The goal of this process is to update similar paths from the situation template based on the paths resulted from the decision tree. In order to do so, a set of similar paths including a similarity score in each pair is built as the result of the Algorithm \ref{al_branches}. Each item in this set represents a pair of similar circumstances that a specific situation can happen and how much similar they are.
\begin{algorithm}[!h]
	\KwIn{dtdnf (a labeled decision tree DNF), stdnf (a labeled situation template DNF) }
	\KwOut{The array of similarities between paths} 
	
	$stpArray \leftarrow stdnf$ paths 
	
	$dtpArray \leftarrow dtdnf$ paths
	
	$similarArray \leftarrow [\:]$
	
	\ForEach{$dtp$ in $dtpArray$}
	{
		$max = 0$
		
		$match = null$
		
		\ForEach{$stp$ in $stpArray$}
		{
			$similarity = similarity-computer(stp,dtp)$ //which is describe in Algorithm \ref{al_similarity}
			
			\If{$similarity >= 0.6$ and $similarity > max$}
			{
				$max = similarity$
				
				$match = stp$
			}
		}
		remove $match$ from $stpArray$
		
		add $[match,dtp,max]$ to $similarArray$
	}
	$return \:\: similarArray$
	\caption{{\bf Making the similarity set} \label{al_branches}}
	
\end{algorithm}
\begin{algorithm}[!h]
	\KwIn{dtp (a path from decision rree DNF), stp (a path from situation template DNF) }
	\KwOut{a similarity score between 0 and 1} 
	
	$stpLeafs \leftarrow stp$ leaves of the situation template DNF
	
	$dtpLeafs \leftarrow dtp$ leaves of the decision tree DNF
	
	$count \leftarrow 0$ 
	$length \leftarrow max(stpLeafs.length,dtpLeafs.length) $
	\ForEach{$snode$ in $stpLeafs$}
	{
		\ForEach{$dnode$ in $dtpLeafs$}
		{
			\If{$snode$ contains Treshold and $dnode$ contains Treshold and $snode$ compartor $=$ $dnode$ comparator}
			{
				\If{$snode.SensorType$ $=$ $dnode.SensorType$ $\&\&$ [$snode$ treshold - $dnode$ treshold] $<=$ 0.25}
				{
					$count = count + 1;$ \: $break;$
				}
			}
			\Else(if\:$snode$ doesn't contain Treshold and $dnode$ doesn't contain Treshold)
			{
				\If{$snode.SensorType$ $=$ $dnode.SensorType$ $\&\&$ $snode.Value$ $=$ $dnode.Value$}
				{
					$count = count + 1;$ \: $break;$
				}    
			}
		}
	}
	$return$ $count \div length$
	\caption{{\bf Similarity Computer Algorithm} \label{al_similarity}}
\end{algorithm}

The algorithm scores each pair of input paths based on their condition nodes and the value of their threshold. Condition nodes can be similar only if they have an identical sensor type and logical operator. There is also a range on which the thresholds can differ. Finally, each pair of paths will be added to the similarity set if and only if the similarity score between them is higher than 0.6 out of 1.

\begin{equation}
train-data = \{(x,y): x= input, y=label\}
\end{equation}

\begin{equation}
Card(A) = \mid\{ (a,b)| (a,b)\in train-data \wedge b="A" \}\mid
\end{equation}

\begin{equation}
%Conf(A) = \sum\nolimits_{(path,conf,card)\in paths(A)}{conf} \div \mid paths(A)\mid
Conf(A) = min\{x_{purity} : x_{purity} \in A(purities) \}
\end{equation}

To be cautious about making changes in situation templates, we define path reliability and label reliability variable that form our restrictions on integrating DNF paths with each other.
The path reliability is computed as the Formula \ref{formula:path-reliability} indicating if the path is reliable or not.
For computing the label reliability measure, label confidence and label cardinality are defined. The label confidence equals the minimum purity of paths in the DNF tree corresponded to the label, and the label cardinality indicates the number of training data holding that label. Finally, the label reliability is determined as shown in Formula \ref{formula:label-reliability}.
	\begin{equation}
	\label{formula:path-reliability}
	Reliability(path,purity,cardinality) = \left\{
	\begin{array}{@{}l@{\thinspace}l}
	0  &: purity < 0.65  \vee  cardinality < 10 \\
	1  &: o.w.\\
	
	\end{array}
	\right.
	\end{equation}

%\begin{strip}
	\begin{equation}
	\hspace*{-0.5cm}
	\label{formula:label-reliability}
	Label-Reliability(A) = \left\{
	\begin{array}{@{}l@{\thinspace}l}
	0  &: Conf(A) < 0.8  \vee  Card(A) < 100 \\
	1  &: o.w.\\
	\end{array}
	\right.
	\end{equation}
%\end{strip}

The ultimate changes on the situation template DNFs are
\begin{enumerate*}
	\item Adding a new path
	\item Removing an old path
	\item Updating a threshold value in an existing path 
\end{enumerate*}.
The first and the second change is applied by considering the whole items of the similarity set with a specific label. Conversely, the third change is based on each pair in this set.

\begin{figure}[!h]
	\centering
	\includegraphics[scale=0.14]{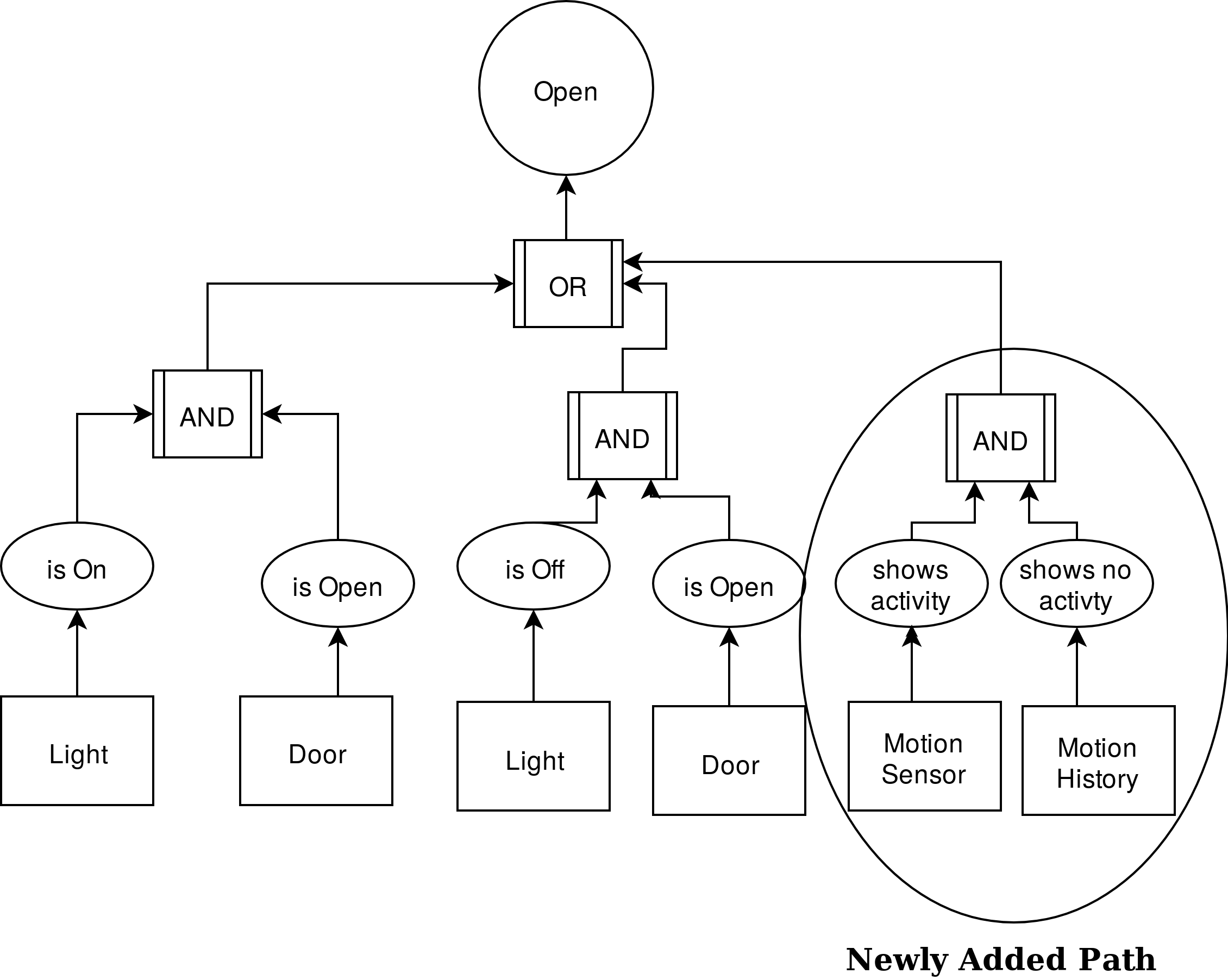}
	\caption{Adding new paths to situation template}
	\label{fig:update_add_path}
\end{figure}

Adding a new path to the situation template's DNF tree occurs when a reliable path from the decision tree does not match to any existing one in the situation template DNF. In this case, the situation template DNF is extended by the new path. A sample result of such a process is shown in Figure \ref{fig:update_add_path}.

Moreover, an old path is removed when a situation template's path is missing from any pairs of the similarity set. However, the path will only be deleted if the corresponded situation template's label is reliable in the decision tree. As the decision tree paths are learned by experiencing events in the environment, there may be some circumstances that occur rarely and will not be reflected on the decision tree. However, these circumstances could be considered by the expert in the situation templates. In order to prevent the removal of such paths, we allow the designers to mark these paths with a "rare flag". As a result, a rare flagged path will never be removed, no matter if it is absent from the similarity set or not.
Removing unmatched paths decreases the execution latency and increases the accuracy of the recognition method. The accuracy is improved as it can reduce the erroneous recognition and prevent misinterpretation of data. Removing a path can also occur when a sensor is corrupted either by sending a fixed or random false data to the processor unit.

If the initial situation templates are designed carefully, the threshold update will be more frequent than the previous modifications. Even if the experts do well on modeling, determining the thresholds of condition nodes is barely possible without analyzing the specific environment for a long period. For instance, some may read books in dimmer light than others do, or some may prefer different temperature thresholds for air conditioning system to function. Therefore, thresholds are not accurately predictable before running the system for a while and are not evident at the initial step of the situation modeling. Thus, this modification is the one which mainly increases the accuracy of situation recognition and personalizes the models for environments.
Threshold modification is done according to each pair of the similarity set which contains a reliable path from the decision tree DNF. In this process, the thresholds of the situation template DNF's path's condition nodes are updated according to the condition nodes in the other pair. For instance, this process can change the conditional comparison $LIGHT-SENSOR-1 > 0.4$ to $LIGHT-SENSOR-1 > 0.8$ in a situation template DNF's path.

At the end of the merging process, the updated situation template DNFs are transformed back to their initial format and replace the existing templates in the repository.
As the final word, if the user changes its behavior or some new sensors are added to the system, a refreshing in machine learning data is required. As machine learning algorithms are built on top of the data provided to them, they normally reflect the longest term conditions and do not reflect new changes at a good pace. Therefore, the user should be able to refresh the machine learning training data. This way, a fresh decision tree is made from new sensors' images to enable reflects on changes.

\section{Evaluation}
\label{sec:evaluation}
We evaluate our approach through a simulation process. In this experiment, a single company consisted of three different rooms is simulated. In each room, there exist some sensors whose types are available in Table \ref{t:sensors}. The smart environment caches the values of sensors every 1 second to enable the detection of changes. The existence of sensors in each room is also available in Table \ref{t:sensors_room}.
\begin{table}[!h]
	\small
	\begin{tabular}{cccc}
		Sensors       & Output    & Description                                    & {wrong outputs on}\\
		Motion Sensor & \{0,1\}   & detect human motion & {not moving humans}\\
		Boolean light & \{0,1\}   & detects the light status& -\\
		Fuzzy light   & {[}0,1{]} & detects the light status& -\\
		Noise Sensor  & {[}0,1{]} & detects room's noise& {outside noises}\\
		TV Sensor     & \{0,1\}   & detects the tv status& -        
	\end{tabular}
	\caption{The types of sensors}
	\label{t:sensors}
\end{table}
\begin{table}[!h]
%	\hspace*{-2.5cm}
	\begin{tabular}{llllll}
		Rooms/Sensors   & {Motion sensor} & {Boolean light} & {Fuzzy light} & {Noise sensor} & {Television sensor} \\
		{Working Room}    & 1             & 0             & 1           & 1            & 1         \\
		{Management Room} & 1             & 0             & 1           & 0            & 0         \\
		Rest Room       & 1             & 1             & 0           & 0            & 0        
	\end{tabular}
	\caption{List of sensors in each room}
	\label{t:sensors_room}
\end{table}

After implanting sensors, five possible situations that the system is supposed to catch is defined. The situations and their description are discussed in Table \ref{t:situations}.
\begin{table}[]
	\begin{tabular}{ll}
		Sensors   & Description                                     \\ \hline
		Opening   & The company is opened (It was previously closed) \\
		Closing   & The company is Closed (It was previously open)  \\
		Working   & the manager or the staff are working            \\
		Educating & An educating session is active in company      
	\end{tabular}
	\caption{Situations to be detected by smart environment}
	\label{t:situations}
\end{table}

We generate random data from these sensors and manually label them to be used as our training and test data. The generated random data is double checked to be feasible sensors' images at two points. First, the initial images are tested by a rule engine to verify the possibility of happening. Secondly, the labeled data are checked by some simple rules to avoid human mistakes. However, we have left 2\% of the errors untouched so that it can reflect the erroneous user feedback in real cases.
Furthermore, we split our generated data into seven parts. One part is considered as our test data and is not provided to our machine learning unit. Each other part represents a plot of time in which the system is working and the user has given feedback on the recognized situations and involves 200-250 sensors' images. This consideration is mandatory because it is unrealistic assumption that all sensors' images are available at the initial step. Thus, we simulate the time by running our algorithm six times on while adding one part to the training data each time assuming they were generated increasingly though time.

As experts and their initial situation models play an essential role in the outcome of our approach, we evaluate our method two times by defining two different sets of initial situation templates. The first set would contain good starting models that are highly accurate. On the other hand, the second set is poorly designed and is used to test how much our approach relies on the starting knowledge and how much it takes for the decision tree to affect the situation models.

\begin{figure*}[!htbp]
%	\hspace*{-3cm}
	\begin{subfigure}{1\linewidth}
		\includegraphics[width=.3\linewidth]{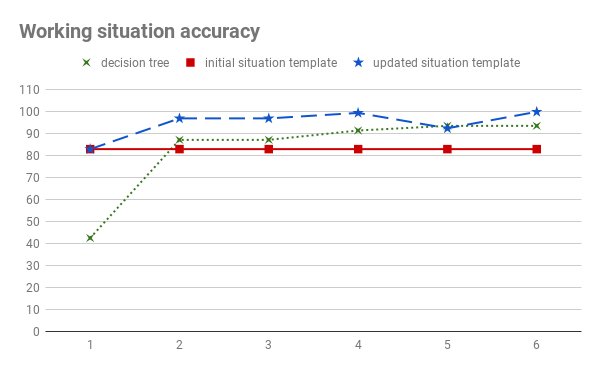}\hfill
		\includegraphics[width=.3\linewidth]{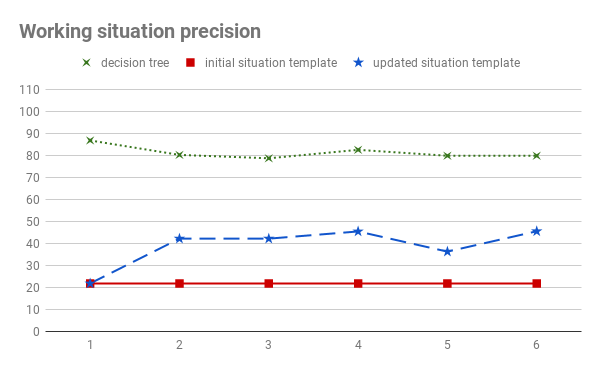}\hfill
		\includegraphics[width=.3\linewidth]{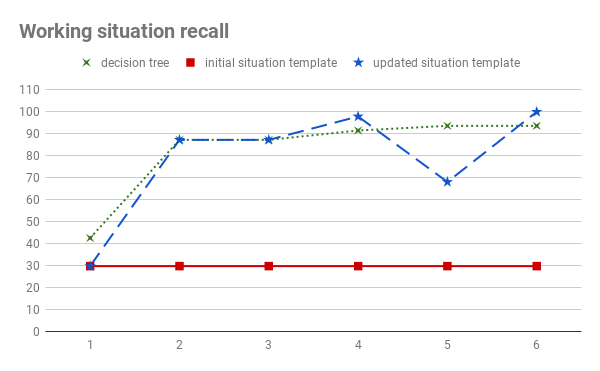}
		\caption{Working situation evaluation}
	\end{subfigure}\par\medskip
%	\hspace*{-3cm}
	\begin{subfigure}{1\linewidth}
		\includegraphics[width=.3\linewidth]{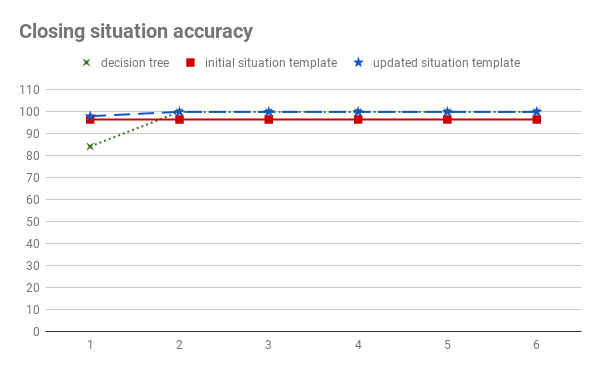}\hfill
		\includegraphics[width=.3\linewidth]{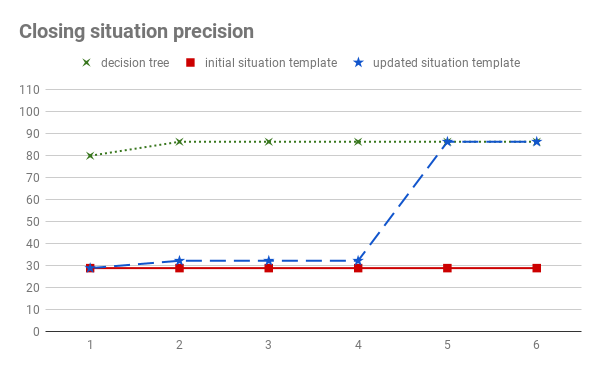}\hfill
		\includegraphics[width=.3\linewidth]{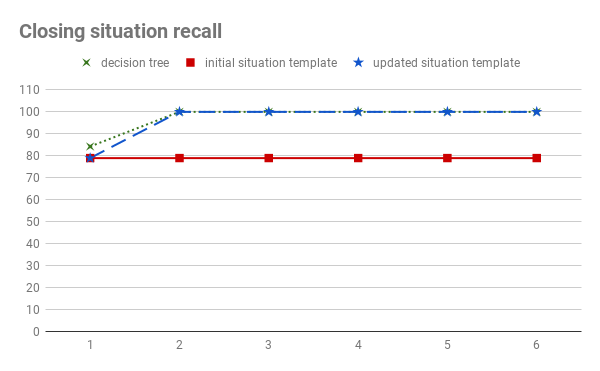}
		\caption{Closing situation evaluation}
	\end{subfigure}\par\medskip
%	\hspace*{-3cm}
	\begin{subfigure}{1\linewidth}
		\includegraphics[width=.3\linewidth]{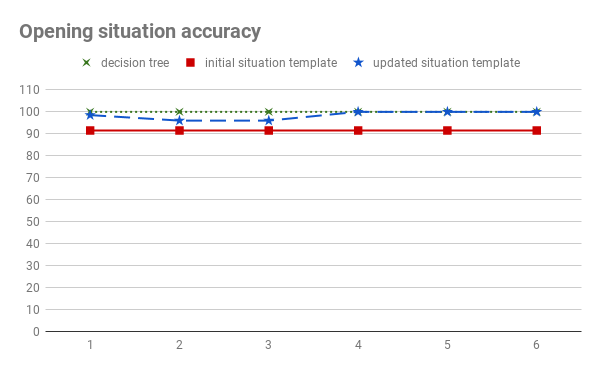}\hfill
		\includegraphics[width=.3\linewidth]{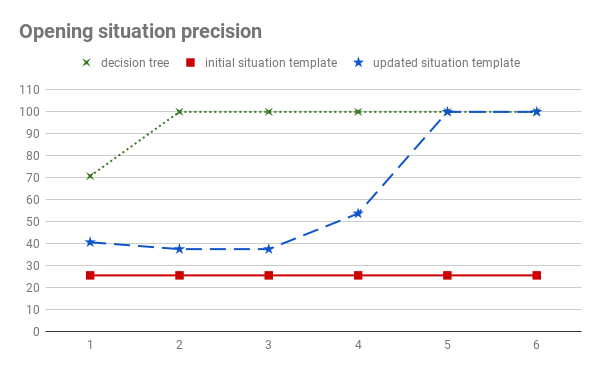}\hfill
		\includegraphics[width=.3\linewidth]{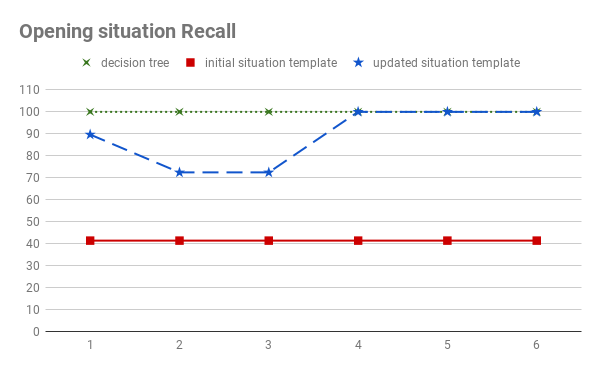}
		\caption{Opening situation evaluation}
	\end{subfigure}
%	\hspace*{-3cm}
	\begin{subfigure}{1\linewidth}
		\includegraphics[width=.3\linewidth]{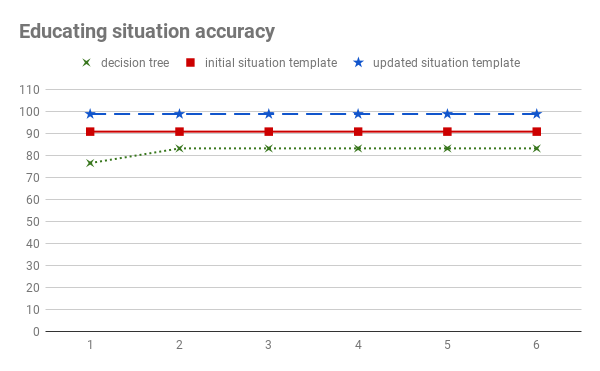}\hfill
		\includegraphics[width=.3\linewidth]{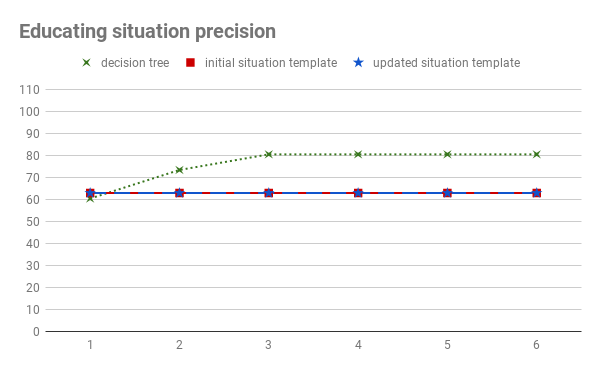}\hfill
		\includegraphics[width=.3\linewidth]{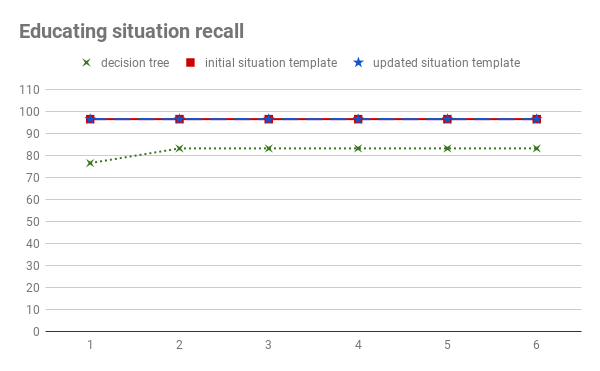}
		\caption{Educating situation evaluation}
	\end{subfigure}
	\caption{Bad start evaluation (accuracy, precision, recall)}
	\label{fig:bad_start}
\end{figure*}

\begin{figure*}[!htbp]
%	\hspace*{-3cm}
	\begin{subfigure}{1\linewidth}
		\includegraphics[width=.3\linewidth]{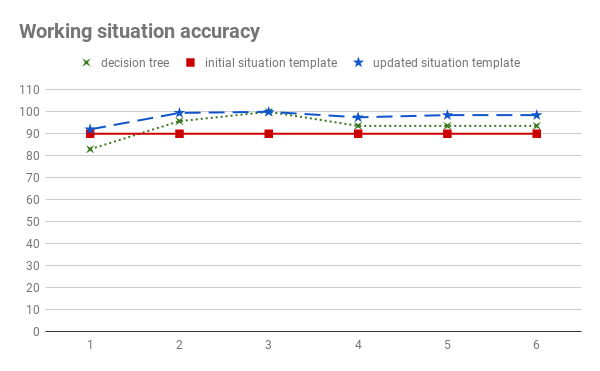}\hfill
		\includegraphics[width=.3\linewidth]{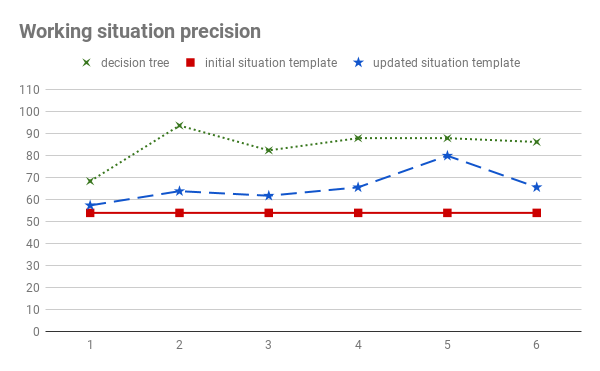}\hfill
		\includegraphics[width=.3\linewidth]{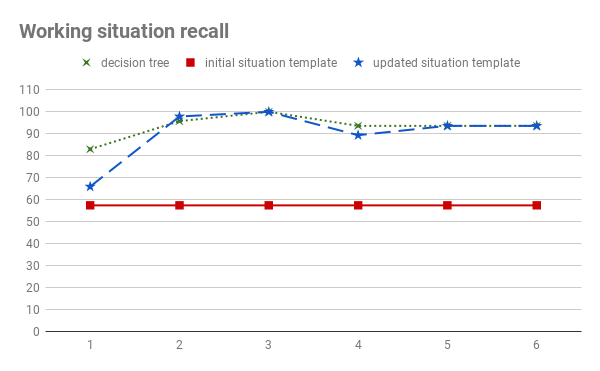}
		\caption{Working situation evaluation}
	\end{subfigure}\par\medskip
%	\hspace*{-3cm}
	\begin{subfigure}{1\linewidth}
		\includegraphics[width=.3\linewidth]{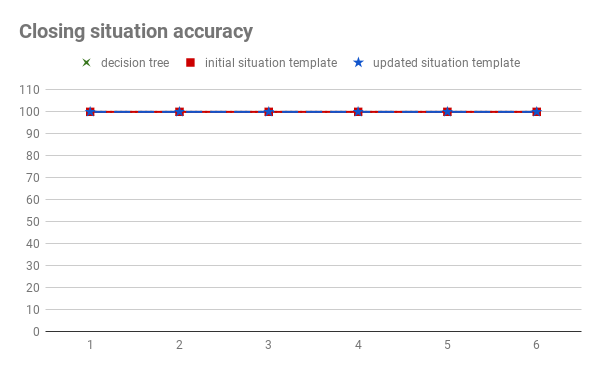}\hfill
		\includegraphics[width=.3\linewidth]{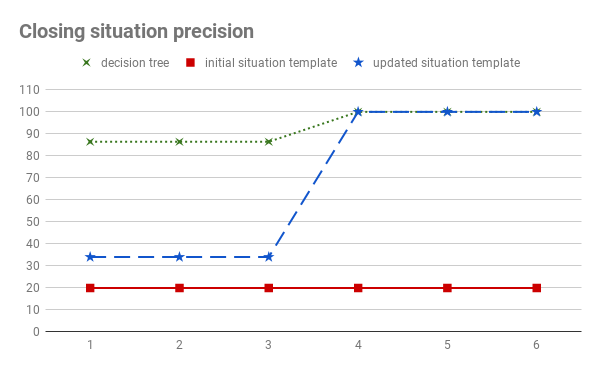}\hfill
		\includegraphics[width=.3\linewidth]{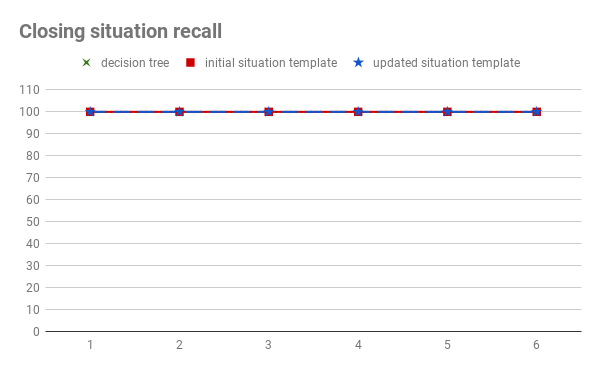}
		\caption{Closing situation evaluation}
	\end{subfigure}\par\medskip
%	\hspace*{-3cm}
	\begin{subfigure}{1\linewidth}
		\includegraphics[width=.3\linewidth]{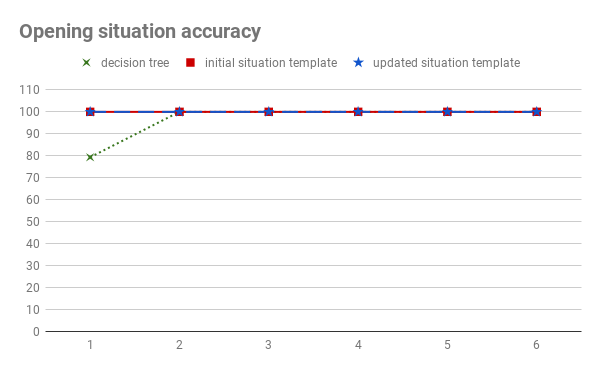}\hfill
		\includegraphics[width=.3\linewidth]{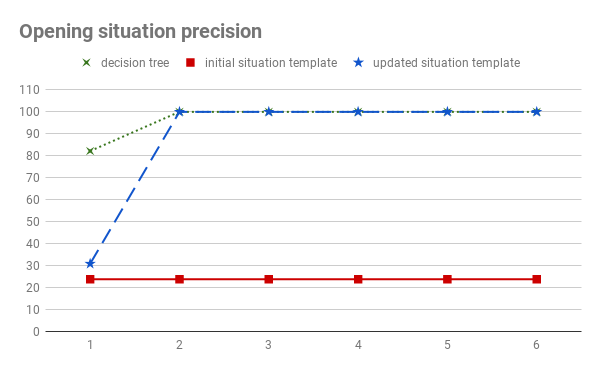}\hfill
		\includegraphics[width=.3\linewidth]{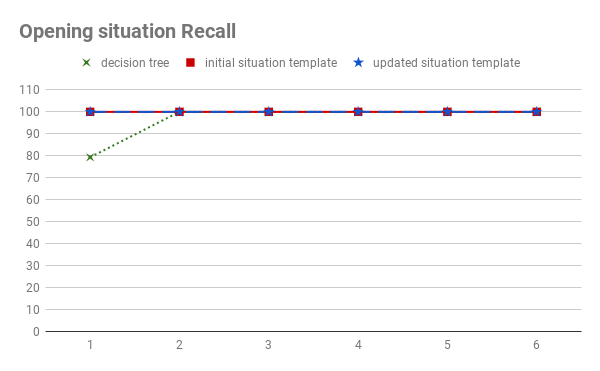}
		\caption{Opening situation evaluation}
	\end{subfigure}
%	\hspace*{-3cm}
	\begin{subfigure}{1\linewidth}
		\includegraphics[width=.3\linewidth]{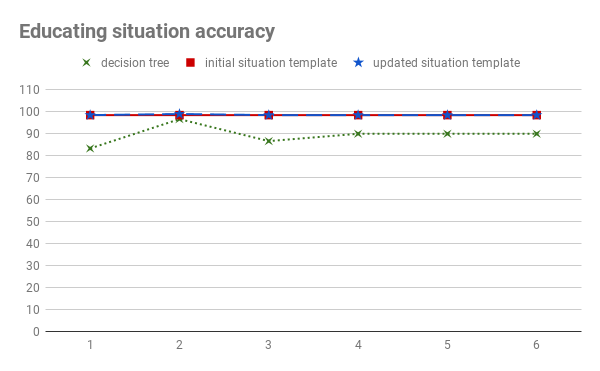}\hfill
		\includegraphics[width=.3\linewidth]{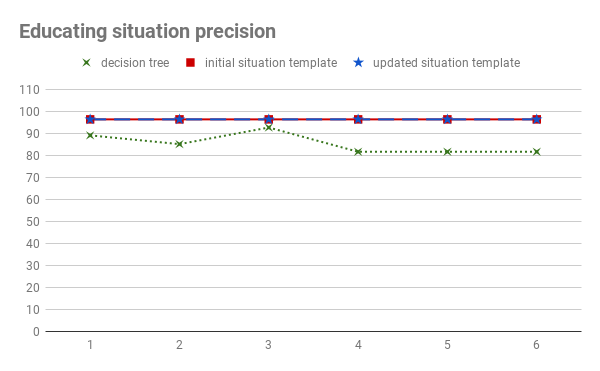}\hfill
		\includegraphics[width=.3\linewidth]{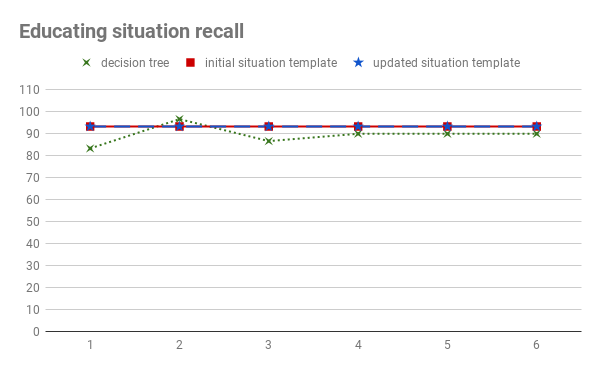}
		\caption{Educating situation evaluation}
	\end{subfigure}
	\caption{Good start evaluation (accuracy, precision, recall)}
	\label{fig:good_start}
\end{figure*}

Finally, we have used the test data to evaluate initial situation templates, updated situation templates and the decision tree accuracy in each step. The result of our evaluation of bad starting models is plotted in Figure \ref{fig:bad_start} while the good start evaluation is depicted in Figure \ref{fig:good_start}. In all the charts the solid line with squares represents the initial situation templates, the dashed line with stars represents the updated templates and the dashed line with cross marks indicates the decision tree. In each part of the figures, the left one shows the accuracy comparison, the centered one represents the precision comparison and the right one shows the differences in recall metric.

The reason that the precision of the decision tree is higher than other two approaches in many cases is that the initial human knowledge had considered general rules that happen to identify the circumstances that the situation is happening while resulting in so much other false identifications that the situation is not occurring. Moreover, there are some points that the updated situation templates are not reflecting the decision tree changes immediately. It has two reasons; First, the machine learning unit has decreased in accuracy and the changes were not reliable at the moment; Second, the cardinality of the paths from the decision tree is not adequate to convince the merging mechanism to integrate them with the situation templates.

The results show that our approach's accuracy, precision, and recall at worst case is equal to the initial situation templates. However, we have a superior in all the metrics in many cases. They also show that our approach does not change on the basis of an inaccurate decision tree in comparison with the initial templates. For instance, the educating situation accuracy, precision, and recall are not changed even if the data of the decision tree is increasing, but its power is not.
An important factor in running our algorithm is to pay special attention in specifying thresholds of reliability as defined in the previous Section. Otherwise, the result may not hold as they have in our case.

\section{Conclusion}
\label{sec:conclusion}
In this paper, we presented a hybrid-learning method to offer situation awareness utilizing situation templates and decision trees. We tested our method against a growing collection of data to simulate the real world example of smart environment's applications. The results showed that our approach guarantees better accuracy in every step of integration in comparison with the initial situation templates. This work can be extended by considering more factors of the real environments in the decision-making process and investigating a proper feedback gathering mechanism that allows the users to contribute in the excellence of the system while not being frequently disturbed. Our defined thresholds for enabling integration and the measure of reliability is also possible to become better; It can be researched to evaluate how the parameters should be computed based on the total number of available data or other critical factors. Also, crowd-sourcing can help the machine learning part to make better decisions by comparing different smart environments to each other. Automatic detection of new events and situations can also be considered using clustering methods while observing the user sequence of actions. Also, as security is of great importance in such systems dealing with plenty of data and the risk of automation, the system should be extended to consider some automation policies and situation security in detection and execution, a feasible solution to this challenge is by considering game theory approaches. However, the challenge of security is mostly considered in the main framework instead of each part of it. 

%\clearpage

\bibliographystyle{IEEEtran}
\bibliography{ref} 

\begin{thebibliography}{10}
\providecommand{\url}[1]{#1}
\csname url@samestyle\endcsname
\providecommand{\newblock}{\relax}
\providecommand{\bibinfo}[2]{#2}
\providecommand{\BIBentrySTDinterwordspacing}{\spaceskip=0pt\relax}
\providecommand{\BIBentryALTinterwordstretchfactor}{4}
\providecommand{\BIBentryALTinterwordspacing}{\spaceskip=\fontdimen2\font plus
\BIBentryALTinterwordstretchfactor\fontdimen3\font minus
  \fontdimen4\font\relax}
\providecommand{\BIBforeignlanguage}[2]{{%
\expandafter\ifx\csname l@#1\endcsname\relax
\typeout{** WARNING: IEEEtran.bst: No hyphenation pattern has been}%
\typeout{** loaded for the language `#1'. Using the pattern for}%
\typeout{** the default language instead.}%
\else
\language=\csname l@#1\endcsname
\fi
#2}}
\providecommand{\BIBdecl}{\relax}
\BIBdecl

\bibitem{mukhopadhyay2014internet}
S.~C. Mukhopadhyay and N.~K. Suryadevara, ``Internet of things: Challenges and
  opportunities,'' in \emph{Internet of Things}.\hskip 1em plus 0.5em minus
  0.4em\relax Springer, 2014, pp. 1--17.

\bibitem{matheus2003core}
C.~J. Matheus, M.~M. Kokar, and K.~Baclawski, ``A core ontology for situation
  awareness,'' in \emph{Proceedings of the Sixth International Conference on
  Information Fusion}, vol.~1, 2003, pp. 545--552.

\bibitem{d2015multi}
G.~D’Aniello, V.~Loia, and F.~Orciuoli, ``A multi-agent fuzzy consensus model
  in a situation awareness framework,'' \emph{Applied Soft Computing}, vol.~30,
  pp. 430--440, 2015.

\bibitem{dey2001understanding}
A.~K. Dey, ``Understanding and using context,'' \emph{Personal and ubiquitous
  computing}, vol.~5, no.~1, pp. 4--7, 2001.

\bibitem{abowd1999towards}
G.~D. Abowd, A.~K. Dey, P.~J. Brown, N.~Davies, M.~Smith, and P.~Steggles,
  ``Towards a better understanding of context and context-awareness,'' in
  \emph{International symposium on handheld and ubiquitous computing}.\hskip
  1em plus 0.5em minus 0.4em\relax Springer, 1999, pp. 304--307.

\bibitem{al2013context}
S.~Al-Sultan, A.~H. Al-Bayatti, and H.~Zedan, ``Context-aware driver behavior
  detection system in intelligent transportation systems,'' \emph{IEEE
  transactions on vehicular technology}, vol.~62, no.~9, pp. 4264--4275, 2013.

\bibitem{wan2014context}
J.~Wan, D.~Zhang, S.~Zhao, L.~T. Yang, and J.~Lloret, ``Context-aware vehicular
  cyber-physical systems with cloud support: architecture, challenges, and
  solutions,'' \emph{IEEE Communications Magazine}, vol.~52, no.~8, pp.
  106--113, 2014.

\bibitem{santa2009sharing}
J.~Santa and A.~F. Gomez-Skarmeta, ``Sharing context-aware road and safety
  information,'' \emph{IEEE Pervasive Computing}, vol.~8, no.~3, pp. 58--65,
  2009.

\bibitem{ferreira2018context}
J.~C. Ferreira, H.~Silva, J.~A. Afonso, and J.~L. Afonso, ``Context aware
  advisor for public transportation,'' \emph{IAENG International Journal of
  Computer Science}, no.~1, pp. 74--81, 2018.

\bibitem{al2018small}
F.~Al-Turjman, E.~Ever, and H.~Zahmatkesh, ``Small cells in the forthcoming
  5g/iot: Traffic modeling and deployment overview,'' in \emph{Smart Things and
  Femtocells}.\hskip 1em plus 0.5em minus 0.4em\relax CRC Press, 2018, pp.
  17--82.

\bibitem{urbieta2017adaptive}
A.~Urbieta, A.~Gonz{\'a}lez-Beltr{\'a}n, S.~B. Mokhtar, M.~A. Hossain, and
  L.~Capra, ``Adaptive and context-aware service composition for iot-based
  smart cities,'' \emph{Future Generation Computer Systems}, vol.~76, pp.
  262--274, 2017.

\bibitem{abdellatif2019edge}
A.~A. Abdellatif, A.~Mohamed, C.~F. Chiasserini, M.~Tlili, and A.~Erbad, ``Edge
  computing for smart health: Context-aware approaches, opportunities, and
  challenges,'' \emph{IEEE Network}, 2019.

\bibitem{aborokbah2018adaptive}
M.~M. Aborokbah, S.~Al-Mutairi, A.~K. Sangaiah, and O.~W. Samuel, ``Adaptive
  context aware decision computing paradigm for intensive health care delivery
  in smart cities—a case analysis,'' \emph{Sustainable cities and society},
  vol.~41, pp. 919--924, 2018.

\bibitem{casino2018smart}
F.~Casino, C.~Patsakis, E.~Batista, O.~Postolache,
  A.~Mart{\'\i}nez-Ballest{\'e}, and A.~Solanas, ``Smart healthcare in the iot
  era: A context-aware recommendation example,'' in \emph{2018 International
  Symposium in Sensing and Instrumentation in IoT Era (ISSI)}.\hskip 1em plus
  0.5em minus 0.4em\relax IEEE, 2018, pp. 1--4.

\bibitem{al2018context}
F.~Al-Turjman and S.~Alturjman, ``Context-sensitive access in industrial
  internet of things (iiot) healthcare applications,'' \emph{IEEE Transactions
  on Industrial Informatics}, vol.~14, no.~6, pp. 2736--2744, 2018.

\bibitem{kolbe2017enriching}
N.~Kolbe, A.~Zaslavsky, S.~Kubler, J.~Robert, and Y.~Le~Traon, ``Enriching a
  situation awareness framework for iot with knowledge base and reasoning
  components,'' in \emph{International and Interdisciplinary Conference on
  Modeling and Using Context}.\hskip 1em plus 0.5em minus 0.4em\relax Springer,
  2017, pp. 41--54.

\bibitem{endsley1995measurement}
M.~R. Endsley, ``Measurement of situation awareness in dynamic systems,''
  \emph{Human factors}, vol.~37, no.~1, pp. 65--84, 1995.

\bibitem{ye2012situation}
J.~Ye, S.~Dobson, and S.~McKeever, ``Situation identification techniques in
  pervasive computing: A review,'' \emph{Pervasive and mobile computing},
  vol.~8, no.~1, pp. 36--66, 2012.

\bibitem{hirmer2017situation}
P.~Hirmer, M.~Wieland, H.~Schwarz, B.~Mitschang, U.~Breitenb{\"u}cher, S.~G.
  S{\'a}ez, and F.~Leymann, ``Situation recognition and handling based on
  executing situation templates and situation-aware workflows,''
  \emph{Computing}, vol.~99, no.~2, pp. 163--181, 2017.

\bibitem{yang2016towards}
K.~Yang, J.~Wang, L.~Bao, M.~Ding, J.~Wang, and Y.~Wang, ``Towards future
  situation-awareness: A conceptual middleware framework for opportunistic
  situation identification,'' in \emph{Proceedings of the 12th ACM Symposium on
  QoS and Security for Wireless and Mobile Networks}.\hskip 1em plus 0.5em
  minus 0.4em\relax ACM, 2016, pp. 95--101.

\bibitem{yang2008using}
J.-Y. Yang, J.-S. Wang, and Y.-P. Chen, ``Using acceleration measurements for
  activity recognition: An effective learning algorithm for constructing neural
  classifiers,'' \emph{Pattern recognition letters}, vol.~29, no.~16, pp.
  2213--2220, 2008.

\bibitem{korpipaa2003bayesian}
P.~Korpip{\"a}{\"a}, M.~Koskinen, J.~Peltola, S.-M. M{\"a}kel{\"a}, and
  T.~Sepp{\"a}nen, ``Bayesian approach to sensor-based context awareness,''
  \emph{Personal and Ubiquitous Computing}, vol.~7, no.~2, pp. 113--124, 2003.

\bibitem{alkhomsan2017situation}
M.~N. Alkhomsan, M.~A. Hossain, S.~M.~M. Rahman, and M.~Masud, ``Situation
  awareness in ambient assisted living for smart healthcare,'' \emph{IEEE
  Access}, vol.~5, pp. 20\,716--20\,725, 2017.

\bibitem{lee2016situation}
S.-Y. Lee and F.~J. Lin, ``Situation awareness in a smart home environment,''
  in \emph{Internet of Things (WF-IoT), 2016 IEEE 3rd World Forum on}.\hskip
  1em plus 0.5em minus 0.4em\relax IEEE, 2016, pp. 678--683.

\bibitem{logan2007long}
B.~Logan, J.~Healey, M.~Philipose, E.~M. Tapia, and S.~Intille, ``A long-term
  evaluation of sensing modalities for activity recognition,'' in
  \emph{International conference on Ubiquitous computing}.\hskip 1em plus 0.5em
  minus 0.4em\relax Springer, 2007, pp. 483--500.

\bibitem{damarla2008hidden}
T.~Damarla, ``Hidden markov model as a framework for situational awareness,''
  in \emph{Information Fusion, 2008 11th International Conference on}.\hskip
  1em plus 0.5em minus 0.4em\relax IEEE, 2008, pp. 1--7.

\bibitem{kanda2008will}
T.~Kanda, D.~F. Glas, M.~Shiomi, H.~Ishiguro, and N.~Hagita, ``Who will be the
  customer?: A social robot that anticipates people's behavior from their
  trajectories,'' in \emph{Proceedings of the 10th international conference on
  Ubiquitous computing}.\hskip 1em plus 0.5em minus 0.4em\relax ACM, 2008, pp.
  380--389.

\bibitem{barwise1989situation}
J.~Barwise, \emph{The situation in logic}.\hskip 1em plus 0.5em minus
  0.4em\relax Center for the Study of Language (CSLI), 1989, no.~17.

\bibitem{gruber1993translation}
T.~R. Gruber, ``A translation approach to portable ontology specifications,''
  \emph{Knowledge acquisition}, vol.~5, no.~2, pp. 199--220, 1993.

\bibitem{chen2009semantic}
L.~Chen, C.~Nugent, M.~Mulvenna, D.~Finlay, and X.~Hong, ``Semantic smart
  homes: towards knowledge rich assisted living environments,'' in
  \emph{Intelligent Patient Management}.\hskip 1em plus 0.5em minus 0.4em\relax
  Springer, 2009, pp. 279--296.

\bibitem{ghimire2017iot}
S.~Ghimire, F.~Luis-Ferreira, T.~Nodehi, and R.~Jardim-Goncalves, ``Iot based
  situational awareness framework for real-time project management,''
  \emph{International Journal of Computer Integrated Manufacturing}, vol.~30,
  no.~1, pp. 74--83, 2017.

\bibitem{pearson2016generic}
R.~Pearson, M.~P. Donnelly, J.~Liu, and L.~Galway, ``Generic application driven
  situation awareness via ontological situation recognition,'' in
  \emph{Cognitive Methods in Situation Awareness and Decision Support
  (CogSIMA), 2016 IEEE International Multi-Disciplinary Conference on}.\hskip
  1em plus 0.5em minus 0.4em\relax IEEE, 2016, pp. 131--137.

\bibitem{machado2017reactive}
A.~Machado, V.~Maran, I.~Augustin, L.~K. Wives, and J.~P.~M. de~Oliveira,
  ``Reactive, proactive, and extensible situation-awareness in ambient assisted
  living,'' \emph{Expert Systems with Applications}, vol.~76, pp. 21--35, 2017.

\bibitem{mormul2017situation}
M.~Mormul, P.~Hirmer, M.~Wieland, and B.~Mitschang, ``Situation model as
  interface between situation recognition and situation-aware applications,''
  \emph{Computer Science-Research and Development}, vol.~32, no. 3-4, pp.
  331--342, 2017.

\bibitem{hirmer2015sitrs}
P.~Hirmer, M.~Wieland, H.~Schwarz, B.~Mitschang, U.~Breitenb{\"u}cher, and
  F.~Leymann, ``Sitrs-a situation recognition service based on modeling and
  executing situation templates,'' in \emph{Proceedings of the 9th symposium
  and summer school on service-oriented computing}, 2015, pp. 113--127.

\bibitem{da2016sitrs}
A.~C.~F. da~Silva, P.~Hirmer, M.~Wieland, and B.~Mitschang, ``Sitrs xt-towards
  near real time situation recognition,'' \emph{Journal of Information and Data
  Management}, vol.~7, no.~1, p.~4, 2016.

\bibitem{cook2009ambient}
D.~J. Cook, J.~C. Augusto, and V.~R. Jakkula, ``Ambient intelligence:
  Technologies, applications, and opportunities,'' \emph{Pervasive and Mobile
  Computing}, vol.~5, no.~4, pp. 277--298, 2009.

\bibitem{ranganathan2004reasoning}
A.~Ranganathan, J.~Al-Muhtadi, and R.~H. Campbell, ``Reasoning about uncertain
  contexts in pervasive computing environments,'' \emph{IEEE Pervasive
  computing}, vol.~3, no.~2, pp. 62--70, 2004.

\bibitem{ke2018automatic}
M.~Ke, B.~Zhu, J.~Zhao, and W.~Deng, ``Automatic drive train management system
  for 4wd vehicle based on road situation identification,'' SAE Technical
  Paper, Tech. Rep., 2018.

\bibitem{zweigle2009supervised}
O.~Zweigle, K.~H{\"a}ussermann, U.-P. K{\"a}ppeler, and P.~Levi, ``Supervised
  learning algorithm for automatic adaption of situation templates using
  uncertain data,'' in \emph{Proceedings of the 2nd International Conference on
  Interaction Sciences: Information Technology, Culture and Human}.\hskip 1em
  plus 0.5em minus 0.4em\relax ACM, 2009, pp. 197--200.

\bibitem{yuan2014context}
B.~Yuan and J.~Herbert, ``Context-aware hybrid reasoning framework for
  pervasive healthcare,'' \emph{Personal and ubiquitous computing}, vol.~18,
  no.~4, pp. 865--881, 2014.

\bibitem{kim2017augmented}
M.~Kim, H.~Kang, S.~Kwon, Y.~Lee, K.~Kim, and C.~S. Pyo, ``Augmented ontology
  by handshaking with machine learning,'' in \emph{Advanced Communication
  Technology (ICACT), 2017 19th International Conference on}.\hskip 1em plus
  0.5em minus 0.4em\relax IEEE, 2017, pp. 740--743.

\bibitem{alberti2013internet}
A.~M. Alberti and D.~Singh, ``Internet of things: perspectives, challenges and
  opportunities,'' in \emph{International Workshop on Telecommunications (IWT
  2013)}, 2013, pp. 1--6.

\bibitem{sukode2015context}
S.~Sukode, S.~Gite, and H.~Agrawal, ``Context aware framework in iot: a
  survey,'' \emph{International Journal}, vol.~4, no.~1, 2015.

\bibitem{safavian1991survey}
S.~R. Safavian and D.~Landgrebe, ``A survey of decision tree classifier
  methodology,'' \emph{IEEE transactions on systems, man, and cybernetics},
  vol.~21, no.~3, pp. 660--674, 1991.

\bibitem{quinlan1993c4}
J.~R. Quinlan, ``C4. 5: Programming for machine learning,'' \emph{Morgan
  Kauffmann}, vol.~38, p.~48, 1993.

\end{thebibliography}

%\begin{thebibliography}{1}
%
%
%\end{thebibliography}

\end{document}